\documentclass{article}
\usepackage{spconf,amsmath,epsfig}

\title{Simultaneous determination of the dielectric relaxation behavior and soil water
       characteristic curve of undisturbed soil samples}
\twoauthors
  {Norman Wagner\sthanks{The authors grateful acknowledge the German Research Foundation (DFG) for support of the project Wa 2112/2-1.}}
    {Institute of Material Research and Testing\\ at the Bauhaus-University Weimar\\Coudraystr. 4, 99423 Weimar, Germany}
  {Katja Lauer}
    {Institute of Soil Science and Soil Conservation\\ Justus-Liebig-University Giessen\\ 35392 Giessen, Germany}

\begin{document}
%
\maketitle
\begin{abstract}
The frequency dependence of soil electromagnetic properties contain valuable information of the porous material
due to strong contributions to the dielectric relaxation behavior by interactions between aqueous pore solution
and mineral phases due to interface effects. Soil hydraulic properties such as matric potential are also
influenced by different surface bonding forces due to interface processes.
For this reason, a new analysis methodology was developed, which allows a simultaneous determination of the
soil water characteristic curve and the dielectric relaxation behavior of undisturbed soil samples.
This opens the possibility to systematically analyze coupled hydraulic/dielectric soil properties for
the development of pedotransfer functions to estimate physico-chemical parameters with broadband
HF-EM measurement techniques.
\end{abstract}
\section{Introduction}
\label{sec:intro} Frequency dependent material properties of porous media such
as soil are not only disturbance quantities in applications with high frequency
electromagnetic (HF-EM) techniques (remote sensing, time domain reflectometry,
ground penetrating radar) but also contain valuable information of the porous material
due to strong contributions to the dielectric relaxation behavior by
interactions between aqueous pore solution and mineral phases
\cite{WagScheu2009a, Wagner2011, Kerr2012}. This circumstance opens the possibility to
estimate physico-chemical parameters such as water content, texture, mineralogy
and matric potential with broadband HF-EM measurement techniques. In this
context, a new analysis methodology was developed, which allows the
simultaneous determination of the soil water characteristic curve and the
dielectric relaxation behavior of soil. For assessment of the
approach a set of 25 undisturbed samples are taken from a 80~cm soil profile of
a GPR test site (Taunus/Germany) with coaxial retention cells developed in
Lauer at al. (2012) \cite{Lauer2011a}. The samples were capillary saturated
followed by a step by step de-watering in a pressure plate apparatus as well as
oven drying at 40~$^\circ$C, equilibrated and the frequency dependent HF-EM
material properties were determined in the frequency range from 1~MHz to 5~GHz
with vector network analyzer technique. The dielectric relaxation behavior were
obtained by inverse modeling with a global optimization algorithm based on a
generalized fractional relaxation model according to Wagner et al. (2011)
\cite{Wagner2011}. Selected relaxation parameters are compared with results
determined by means of empirical equations and frequently used mixture models.
\section{Material and Methods}
\label{sec:materials}

Undisturbed soil samples were taken in four depths (0, 30, 50 and 75~cm)
from a typically soil profile developed in the Taunus area,
in the south-eastern part of the Rhenish Massif, Germany (see \cite{Lauer2011a} for details).
In Table \ref{tab:SoilProp} physical, chemical and mineralogical soil properties are summarized.

\subsection{Soil water characteristic curve (SWCC)}
To obtain simultaneously SWCC and frequency dependent HF-EM properties of the soil samples, a
two-port coaxial transmission line cell according to Lauer et al.
\cite{Lauer2011a} was used. The outer diameter of the inner conductor is
16.9~mm, the inner diameter of the outer conductor is 38.8~mm with the total
length of 50~mm. Both conductors are designed with a cutting edge allowing
easier insertion of the cell into the soil without disturbing the natural in
situ soil structure. The taken soil samples were measured as-received, water saturated and stepwise
dewatered by increasing negative pressure
(pF 1.4/1.8/2/2.5/4.2) in a pressure plate apparatus. After each pressure step
the samples were sealed, equilibrated, weighted and the dielectric spectra were
measured. After increasing the pressure to pF 4.2, permittivity measurements
were carried out for saturated and air-dried samples. The in-situ bulk
densities were obtained after drying at 105~$^\circ$C.
\begin{table*}[ht]
\small
  \centering
  \caption{Physical, chemical and mineralogical properties of the investigated soil (see \cite{Lauer2011a}).
      }\label{tab:SoilProp}
\begin{tabular}{l|llll|cccc|r|c}
  \hline
  horizon & sand & silt & clay & organic & vermiculite/ &  illite/kaolinite/ & tecto- & goethite & particle  & cation exchange \\
          &      &      &      &         & smectite    & mixed layer      &  silicates &         & density         &   capacity \\
          & [wt. \%] &  &      &         & [wt. \%]    &                  &            & & [g/cm$^3$]    &  [mmol/100g]\\
  \hline
  \hline
  Ah  & 18.9 & 57.5 & 24.0 & 2.4 & 2.8 / - & 40.9 / 19.1 / 8.1 & 28.4 & 0.7 & 2.62 & 9.02   \\
  Btg & 20.0 & 46.6 & 33.5 & - & 8.7 / 1.8 & 35.5 / 21.8 / 12.7  & 18.5 & 1.1 & 2.65 & 12.51  \\
  2Cg & 43.7 & 31.1 & 25.2 & - & 2.2 / - & 50.7 / 30.9 / 11.3 & 4.9  & - & 2.69 & 11.45 \\
  3Cg & 61.8 & 20.5 & 17.7 & - & 2.8 / - & 51.2 / 27.6 / 9.2 & 8.8 & 0.4 & 2.71 & 11.09 \\
  \hline
\end{tabular}
\end{table*}
\begin{figure*}[ht]
\center
  \includegraphics[scale=0.6]{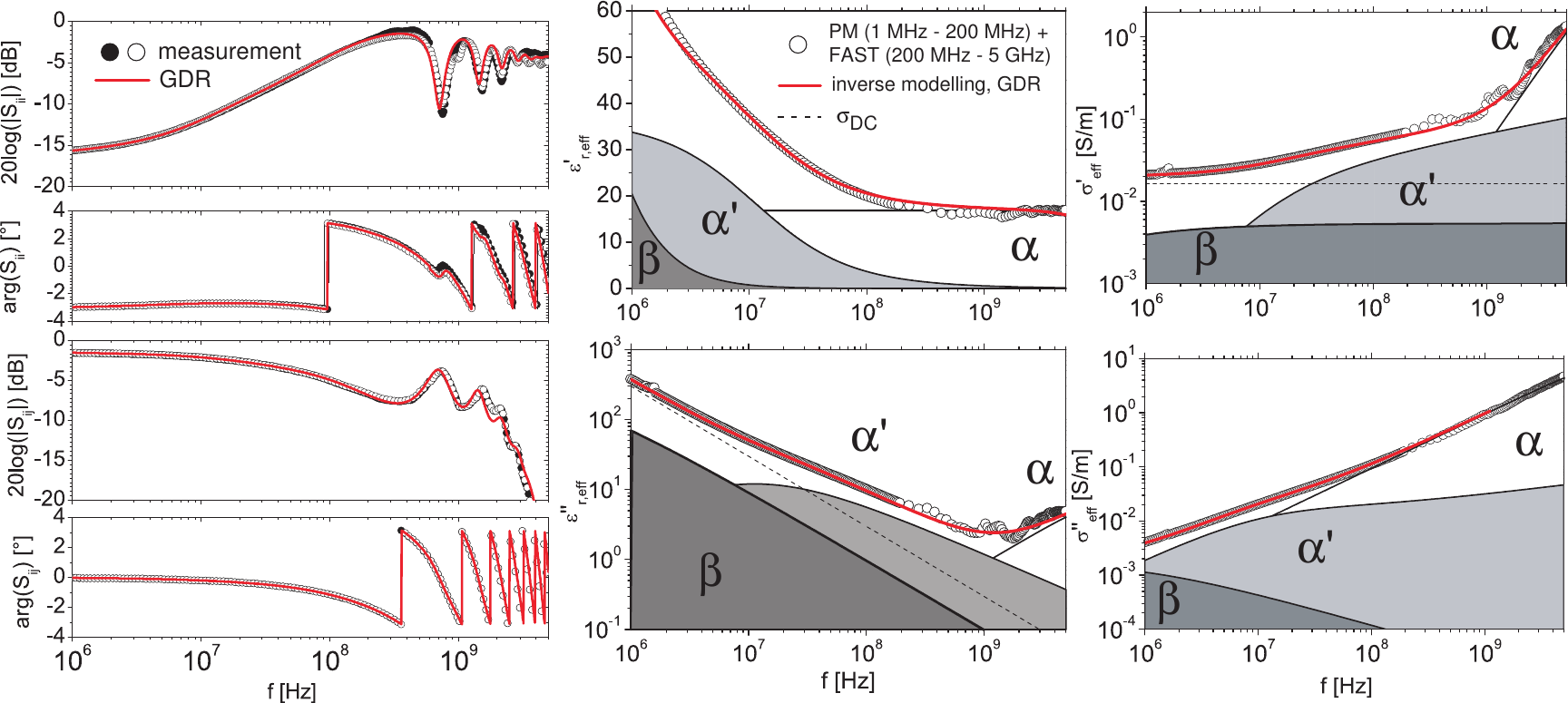}
  \caption{(from left to right) S-Parameter $S_{ij}$, complex effective relative permittivity  $\varepsilon^\star_{\mbox{r,eff}}$
  and complex effective electrical conductivity $\sigma^\star_{\mbox{eff}}$
  as a function of frequency of a sample from the 3Cg horizon at pF 1.8 with $\theta=$~0.29~$m^3m^{-3}$ and $n=0.41$ as well as
  the results of the SCEM-UA optimization (see text for the used terminology).}\label{fig:SoilSpectrum}
\end{figure*}

\subsection{High frequency electromagnetic technique}
\label{sec:HFEMTech}

HF-EM properties were determined within a frequency range from 1~MHz
to 5~GHz at room temperature and atmospheric pressure with
Rohde \& Schwarz ZVR (1~MHz to 4~GHz) and Agilent PNA E8363B (10~MHz to 5~GHz) vector network analyzers.
Full two-port calibration was done by mechanical (Rhode \& Schwarz N - 50 $\Omega$ ZV-Z21)
or electronically (Agilent electronic calibration kit N4691B) calibration standards
(Open, Short, 50 $\Omega$-Match, Through) at the N connector of the high-precision
coaxial cable to the measurement cell. Measurement quantities are the complex
scattering parameters $S_{ij}$ of the full length coaxial line including N to 1~5/8'' EIA
coupling elements at both sample cell ends (Figure \ref{fig:SoilSpectrum}). Complex effective relative permittivity $\varepsilon_{\mbox{r,eff}}$
was calculated by means of Agilent 85071/E materials measurement software
at the frequency range from 200~MHz to 5~GHz.

In addition to Agilent 85071/E materials measurement software, complex
S-parameter values $S_{ij}$ measured with Rohde \& Schwarz ZVR were used to
compute $\varepsilon_{\mbox{r,eff}}$ in the frequency range between 1~MHz to
4~GHz using the following  methods implemented in matlab: classical
Nicholson-Ross-Weir model (NRW), Baker-Jarvis (BJ), BJ-iterative (BJI) and propagation matrix method (PM)
\cite{Lauer2011a, Wagner2011}.
The quasi-analytical methods were compared and validated against inverse
modeling technique according to \cite{Wagner2011} based on a
generalized fractional dielectric relaxation model (GDR):
\begin{eqnarray}\label{eq:GDR}
\varepsilon_{\mbox{r,eff}}^\star-\varepsilon_\infty=
\sum\limits_{k = 1}^N {\frac{{\Delta\varepsilon _k }}{{\left( {j\omega \tau _k} \right)^{\alpha _k} + \left( {j\omega \tau _k}
\right)^{\beta _k} }}} - j\frac{{\sigma _{DC}}}{{\omega \varepsilon _0 }}
\end{eqnarray}
with high frequency limit of permittivity $\varepsilon_\infty$, relaxation
strength $\Delta\varepsilon_k$, relaxation time $\tau_k$ as well as stretching
exponents $0\leq\alpha_k, \beta_k$ of the $k$-th process and apparent direct
current electrical conductivity $\sigma_{DC}$.

The GDR parametrization is performed with a shuffled complex evolution metropolis
algorithm (SCEM-UA) according to Vrugt et al. 2003 \cite{Vrugt2003} assuming
three active relaxation processes in the investigated
frequency-temperature-pressure range (see \cite{WagScheu2009a, Wagner2011}):
one primary $\alpha$-process (main water relaxation) and two secondary
processes caused by solid-water-ion interactions $\alpha'$, $\beta$ (the superposing of relaxation processes due to
adsorbed and hydrated water, counter ion relaxation as well as Maxwell-Wagner effects). The applicability of the HF-EM methodology was
assessed by repeated measurements on a homogeneous dispersive and dielectric as
well as electrical lossy nearly saturated gypsum sample and weak or non
dispersive and low loss common standard materials: air, teflon measured before
and after each pressure step as well as glass, zircon and baddeleyite beads
with air (for details see \cite{Lauer2011a}).
\begin{figure}[ht]
\center
  \includegraphics[scale=0.46]{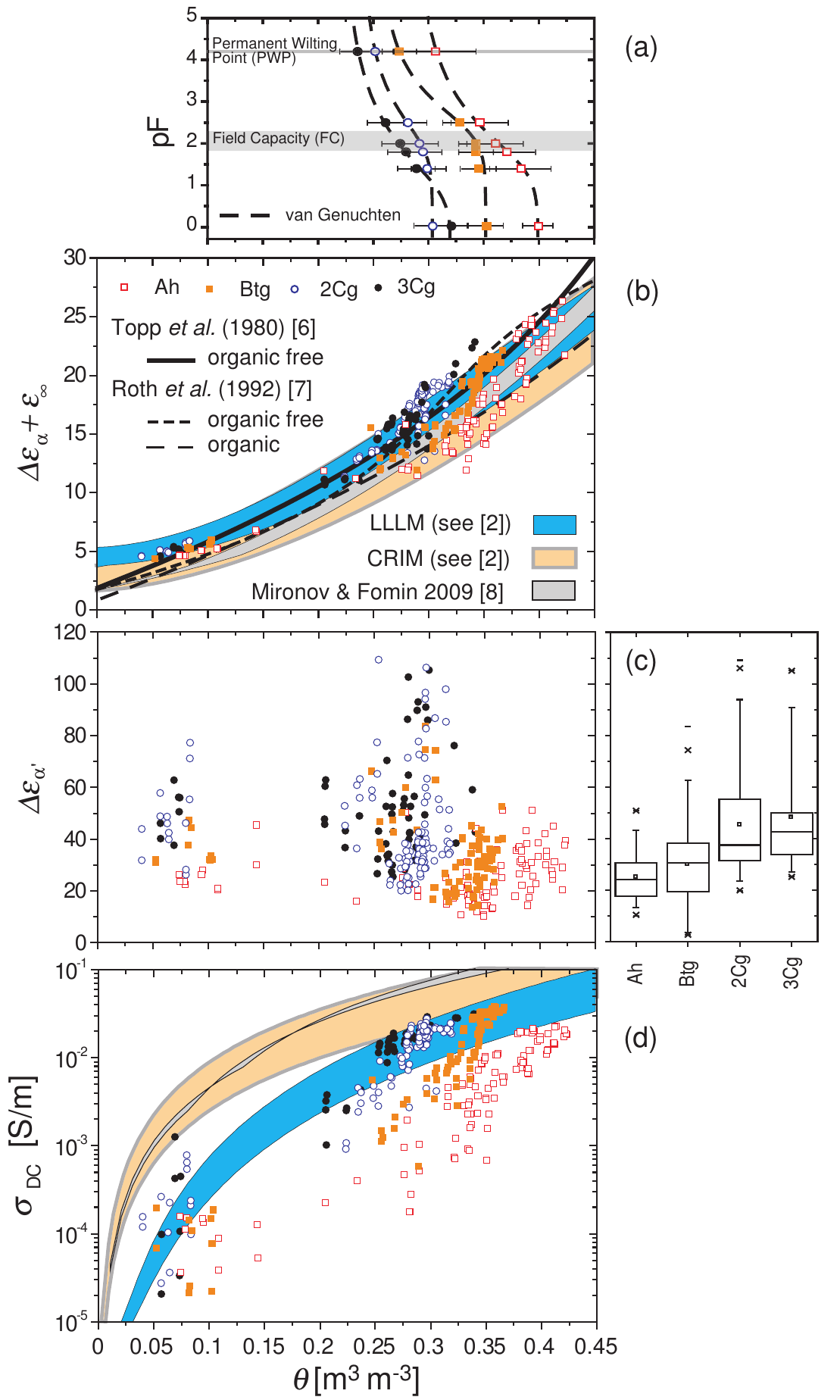}
  \caption{(a) Soil water characteristic curve, (b) high frequency limit of permittivity $\varepsilon_\infty$ and relaxation strength $\Delta\varepsilon_{\alpha}$  of free and immobile pore water, (c) relaxation strength $\Delta\varepsilon_{\alpha'}$ of the $\alpha'$-process as well as box and whiskers plots of  $\Delta\varepsilon_{\alpha'}$ for the appropriate soil horizon and (d) apparent direct current conductivity $\sigma_{DC}$ obtained with inverse modeling as a function of volumetric water content $\theta$ (for details see \cite{Lauer2011a}).}\label{fig:GDRPar}
\end{figure}

\section{Results and Discussion}
\label{sec:results}

In Figure \ref{fig:SoilSpectrum}, the results of the direct inversion algorithms (PM+FAST) and the
inverse modeling technique are represented for a soil sample from 3Cg-horizon at pF 1.8 with $\theta=$~0.29~$m^3m^{-3}$  and $n=0.41$.
Clearly visible in the dielectric spectrum are the $\alpha$- and  $\alpha'$-process with relaxation times of 9~ps or 17~ns,
respectively. In the case of the $\beta$-process only the high frequency tail is
visible because of a relaxation time of 0.8~$\mu$s. Hence, in the frequency
range below 10~MHz, the frequency dependence of the effective complex relative
permittivity was dominated by the $\beta$-process as well as the direct current conductivity contribution.

%

In Figure \ref{fig:GDRPar} the results of the parametrization for all samples
are represented for the dominant  $\alpha$- and  $\alpha'$-process as well as
the apparent direct current conductivity $\sigma_{DC}$ in comparison to the empirical
models according to Topp et al. (1980) \cite{Topp80} and Roth et al. (1992)
\cite{Roth1992}, the semi-empirical generalized refractive mixing dielectric
model (GRMDM) by Mironov et al. (2009) \cite{Mironov2009} as well as the
theoretical mixture equations CRIM (Complex Refractive Index model) and LLLM
(Looyenga-Landau-Lifschitz model) according to \cite{Wagner2011}.

The achieved
mean relative error in the relaxation strength of the $\alpha$- process is
below 1\% in contrast to the $\alpha'$- process with 24 \%. Therefore, the
$\alpha$-process can be related to the volumetric water content, which is also
confirmed by the empirical equations. In the low water content range the Topp
et al. (1980) \cite{Topp80} equation gives better results than
the Roth et al. (1992) \cite{Roth1992} equation
and in the water content range above 0.2~m$^3$m$^{-3}$ vise versa. In Table \ref{tab:RMSE}
appropriate RMSEs are summarized.

GRMDM underestimate the permittivity in the low water content range below
0.2~m$^3$m$^{-3}$ for all soils and above for the soil from the 2Cg and 3Cg
horizon and overestimate the permittivity for the soil from the Ah horizon in
the range below 0.4~m$^3$m$^{-3}$. Hence, the different soil texture, structure
and mineralogy are not able to predict. The relaxation strength of the
$\alpha'$-process shows complicated dependence on volumetric water content.
Especially around 0.30~$m^3m^{-3}$, it rises up to 100 for all samples with
exception of the Ah-horizon.
\begin{table}[t]
\small
  \centering
  \caption{RMSE of the obtained volumetric water content in volume \% from the relaxation strength of the $\alpha$-process with the empirical equations
  for organic free / $^\clubsuit$organic soils.
      }\label{tab:RMSE}
\begin{tabular}{p{3.5cm}|llll}
  \hline
   & Ah & Btg & 2Cg &  3Cg \\\hline\hline
  Topp et al. (1980)  & 4.7 (3.5)$^\clubsuit$ & 2.1 & 2.3 &  2.2 \\
  Roth et al. (1992) & 4.7 (3.4)$^\clubsuit$ & 2.3 & 2.1 & 2.2\\
  \hline
\end{tabular}
\end{table}

Apparent direct current conductivity $\sigma_{DC}$ shows a clear textural
dependence. Moreover, $\sigma_{DC}$ is clearly overestimated with GRMDM due to the
exponent $0.5$ in the underlying CRIM equation and the difficulty to estimate
the conductivity of pore water a priori \cite{Wagner2011}. The theoretical
models CRIM and LLLM are more flexible to characterize the influence of the
pore water conductivity as well as the different physico-chemical soil
properties. However, with CRIM the same overestimation of the direct current
conductivity is observed as in the case of the GRMDM regardless pore water
conductivity is estimated according to the approach in \cite{Wagner2011}.
This suggests to link soil water potential to the pore water conductivity.

The scattering of the relaxation strength of the $\alpha'$-process can be further
attributed to the influence of the non-homo\-geneous structure of the undisturbed samples
in the cell as well as the imposition of the assumed relaxation processes and the difficulty
separating them clearly with the chosen relaxation model in the investigated frequency range.
Nevertheless, the strongest effects in the relaxation strength can be observed for the 2Cg- and
3Cg-horizon with smallest porosities and a permanent wilting point $\theta_{PWP}$=$25\pm2$ and $24\pm2$
respectively. In case of the soil from the Ah-horizon with $\theta_{PWP}$=$31\pm4$  relaxation strength $\Delta\varepsilon_{\alpha'}$ is nearly independent
of water content corresponding with the lowest $\sigma_{DC}$ and indicating the impact of organic matter.
The samples from Btg-horizon with $\theta_{PWP}$=$27\pm2$ as well as the highest clay content and highest amount on swelling clay minerals show an intermediate behavior.

\section{Conclusion}
\label{sec:results}

A new analysis methodology was developed, which
allows a simultaneous determination of the soil water
characteristic curve and the dielectric relaxation behavior of undisturbed
soil samples. For assessment of the approach a set of 25 undisturbed samples from a
soil profile of a GPR test site (Taunus / Germany, \cite{Lauer2011a}) were analyzed
in the frequency range from 1~MHz to 5~GHz with vector network analyzer technique.
The dielectric relaxation behavior was determined by means of inverse modeling  assuming three active relaxation processes:
one primary $\alpha$-process (main water relaxation) and two secondary
processes $\alpha'$, $\beta$ caused by solid-water-ion interactions .
Frequently used empirical equations were related to the free water $\alpha$-process which clearly confirm the great value
to estimate the volumetric water content with this approaches.
GRMDM according to Mironov et al. (2009) \cite{Mironov2009} used at a frequency of 1~GHz is unable to predict textural, mineralogical and structural influences on the permittivity of the $\alpha$-process and clearly overestimate apparent direct current conductivity contribution.
The theoretical Models CRIM and LLLM are more flexible to characterize the influence of the
pore water conductivity as well as the different physico-chemical soil
properties.

The relaxation strength $\Delta\varepsilon_{\alpha'}$  of the
$\alpha'$-process shows a complicated dependence on volumetric water content.
The strongest effects
can be observed for the soil from the basal periglacial slope deposit and bedrock
with smallest porosities. $\Delta\varepsilon_{\alpha'}$ of the Ah-horizon is nearly independent
of water content corresponding with the lowest $\sigma_{DC}$ and indicating the impact of organic matter.
However, the verification and validation of these observations need further systematic analysis of the obtained dataset using relaxation models in combination with mixture equations \cite{WagScheu2009a, Wagner2011} under consideration of the soil structure in the cell in comparison with homogeneous disturbed samples.


\bibliographystyle{IEEEtr}
\bibliography{Literatur20112311}

\end{document}